\documentclass[preprint,showpacs,preprintnumbers,prb]{revtex4}

\usepackage{graphicx}
\usepackage{dcolumn}
\usepackage{bm}

\begin{document}

\title{Photoinduced spin crossover in Fe-picolylamine 
complex: \\ A farinfrared study on single crystals}    

\author{H. Okamura}
  \email[E-mail: ]{okamura@kobe-u.ac.jp}
\author{M.~Matsubara}
\author{T.~Nanba}
\affiliation{Graduate School of Science and Technology, 
Kobe University, Kobe 657-8501, Japan.}

\author{T.~Tayagaki}
 \altaffiliation[Present address: ]{SORST, Japan Science 
and Technology Corporation, Tokyo 113-8656, Japan.}
\author{S.~Mouri}
\author{K.~Tanaka}
\affiliation{Department of Physics, Graduate School 
of Science, Kyoto University, Kyoto 606-8502, Japan. }

\author{Y.~Ikemoto}
\author{T.~Moriwaki}
\author{H.~Kimura}
\affiliation{Japan Synchrotron Radiation Research Institute 
and SPring-8, Sayo 679-5198, Japan.}

\author{G. Juh\'{a}sz}
\altaffiliation[Present address: ]{Department of Chemistry, 
Carnegie Mellon University, Pittsburgh, Pennsylvania 15213.}   
\affiliation{Department of Chemistry, Kyushu University, 
Fukuoka 812-8581, Japan.}

\date{\today}

\begin{abstract}
Farinfrared spectroscopy has been performed on 
[Fe(2-picolylamine)$_3$]Cl$_2$EtOH (Fe-pic) single crystals, 
to probe changes in the molecular vibrations upon 
the photoinduced and temperature-induced spin crossovers.    
Synchrotron radiation has been used as the farinfrared source 
to overcome the strong absorption and the small sizes of 
the samples.  
Absorption lines due to FeN$_6$ cluster vibrations, observed 
below 400~cm$^{-1}$, show strong intensity variations upon the 
crossover due to the deformation of FeN$_6$ between 
high-spin and low-spin states.    
However, they remain almost unchanged between the photo- 
and temperature-induced high-spin states.     
This is in sharp contrast to the lines at 500-700~cm$^{-1}$ 
due to intramolecular vibrations of the picolylamine ligands, 
which show marked variations between the two high-spin states.     
It is concluded that the most important microscopic difference 
between the two high-spin states arises from the ligands, 
which is likely to reflect different states of intermolecular 
bonding between them.

\end{abstract}

                             
\maketitle

[Fe(2-pic)$_3$]Cl$_2$EtOH (2-pic: 2-picolylamine or 
2-aminomethyl pyridine, EtOH: ethanol), referred to as 
the Fe-pic, is one of the Fe(II)-based complexes which 
have recently attracted much interest for exhibiting 
a photoinduced spin crossover.\cite{time,koshihara,SC-review}    
As sketched in Fig.~1(a), an Fe$^{2+}$ ion in Fe-pic is 
located in a nearly octahedral crystal field created 
by the three 2-pic ligands.     
Depending on the magnitude of the crystal field splitting, 
Fe$^{2+}$ takes either total spin $S$=2 (high spin) or $S$=0 
(low spin), as sketched in Fig.1~(b).   Well above 
$T_{1/2} \simeq$ 
118~K, Fe-pic is in the high-spin state (high-temperature 
high-spin state, HTHS).    Upon cooling through $T_{1/2}$, 
Fe-pic undergoes a crossover to the low-spin state 
(low-temperature low-spin state, LTLS).  The width of the 
crossover is about 20~K, as observed in the 
high-spin fraction, $\gamma_{HS}(T)$.     The crossover 
is associated with a $\sim$ 8~\% change in the average 
Fe-N distance.   The resulting electron-lattice coupling 
leads to cooperative interaction among the Fe$^{2+}$ 
ions,\cite{SC-review} as discussed later.     
In LTLS, a photoexcitation can cause a crossover to a 
high-spin state.    
Below 40~K, this photoinduced high-spin state (PIHS) persists 
even after the excitation is turned off.    The half life 
in $\gamma_{HS}$ reaches 160~min at 10~K.\cite{time}     
The development of PIHS involves highly nonlinear responses 
to the photoexcitation, such as an incubation time, a 
threshold intensity, a step-like change of $\gamma_{HS}$ 
with time, and a domain formation.\cite{koshihara}    
These results led to the suggestion that the development of 
PIHS in Fe-pic should be a transition to a novel macroscopic 
phase under photoexcitation, referred to as the 
{\it ``photoinduced phase transition''.}\cite{koshihara}

To examine this suggestion, it is important to compare the 
microscopic nature of PIHS with that of HTHS, and to 
characterize the cooperative interaction among Fe$^{2+}$ 
ions under photoexcitation.      
It was found\cite{taya-PRL,taya-PRB,okamura} that the 
Raman-infrared selection rules in Fe-pic were modified 
between HTHS and PIHS, which suggested a photoinduced 
symmetry lowering in PIHS.      
In contrast, detailed X-ray diffraction (XRD) study of 
Fe-pic\cite{huby} found no significant difference 
in the average crystal structure between HTHS and PIHS.    
In addition, X-ray absorption fine structure (XAFS) of 
Fe-pic showed that the local coordinations of the neighboring 
ions around Fe$^{2+}$ were very similar between HTHS and 
PIHS.\cite{oyanagi}     
Recently, a nuclear resonant inelastic scattering 
(NRIS) experiment was reported on $^{57}$Fe-enriched 
Fe-pic.\cite{juhasz}        
This technique was able to selectively probe the partial 
vibrational density of states for $^{57}$Fe.    
The obtained NRIS spectra 
were very similar between PIHS and HTHS, showing that 
the atomic vibration state of Fe$^{2+}$ was also similar.

In this work, we have measured the farinfrared (FIR) absorption 
spectra of Fe-pic single crystals in the 80-700~cm$^{-1}$ range.   
Unlike the previous midinfrared work,\cite{okamura} this range 
can cover the normal-mode vibrations of the FeN$_6$ cluster.    
To analyze the observed data, the frequencies of molecular 
vibrations are calculated for [Fe(2-pic)$_3$]$^{2+}$ using 
density-functional method.    
Many of the observed absorption lines below 400 cm$^{-1}$ 
are attributed to the FeN$_6$ cluster vibrations, which 
exhibit characteristic intensity changes upon the photo- 
and temperature-induced spin crossovers.    However, they 
are almost unchanged between HTHS and PIHS.    
It is concluded that the microscopic vibrational states of 
the FeN$_6$ cluster is nearly unchanged between PIHS and HTHS, 
and that the deformation of 2-pic ligand should play 
an important role in the development of PIHS.

The single crystals of Fe-pic used in this work were grown by 
the evaporation method.     Plate-shaped samples of approximately 
0.7 $\times$ 0.7 $\times$ 0.1~mm$^3$ were obtained by cleaving 
the crystals, and mounted on a continuous-flow liquid He 
cryostat.     The FIR absorption experiment was 
done using a synchrotron radiation (SR) source and a 
custom-made microscope at the beam line BL43IR, 
SPring-8.\cite{micro}    
The SR source can deliver much higher photon flux density 
to the sample than the usual FIR sources.   Since the 
Fe-pic single crystals had small sizes and strong 
FIR absorption, the use of SR was crucial to 
successfully perform this experiment.  
A black polyethylene filter was used to cut the visible 
component of the SR.    Photo-excitation of the sample 
was made using white light from a tungsten lamp.   A Si 
bolometer was used as a detector, and a Fourier-transform 
interferometer was used to record the spectra.    
The spectral resolution was set to 4~cm$^{-1}$.

Figure~2(a) shows the FIR absorption spectra of Fe-pic 
without photoexcitation at several temperatures across 
$T_{1/2}$.    
Here, the absorption is expressed as the optical density 
(OD), $-\log[I(\nu)/I_0(\nu)]$, where $I(\nu)$ and 
$I_0(\nu)$ are the transmission spectra with and 
without the sample, respectively.     
The detection limit for the weak transmission 
was about OD=2.8 with an accumulation time of 
2~min, and Fig.~2 shows the spectra below OD=2.8 only.     
The lines above 450~cm$^{-1}$ agree well with those 
previously observed, which result from the intramolecular 
vibrations of the 2-pic ligand and 
ethanol.\cite{okamura}    In contrast, many of the lines 
below 450~cm$^{-1}$ are attributed to vibration modes 
of the FeN$_6$ cluster, as discussed later.     
The absorption lines show quite strong intensity 
variations through $T_{1/2}$.   
The detailed temperature dependence of several bands 
and lines, indicated by the labels in Fig.~2, 
are displayed in Fig.~3.     It is seen that the 
variation of the absorption strength occurs over a width 
of about 20~K,\cite{footnote4} which agrees well with 
that of $\gamma_{HS}$.\cite{okamura}     
Namely, the absorption strength closely reflects the 
evolution of electronic configuration at Fe$^{2+}$ and 
the associated deformation in the Fe-pic molecule 
upon the crossover.

Figures~4(a)-(c) present the absorption spectra of Fe-pic 
in LTLS at 9~K, in PIHS at 9~K, and in HTHS at 140~K, 
respectively.\cite{footnote3}    
Here, spectrum (a) was first recorded, then the sample 
was photoexcited with a 2~mW/mm$^2$ power density of white 
light for 5~min.    After turning off the excitation, 
spectrum (b) was measured, then the sample was warmed up 
and spectrum (c) was recorded.   
The spectra above 
450~cm$^{-1}$ agree well with those reported 
previously:\cite{okamura}   The double lines at 530-570~cm$^{-1}$, 
marked by the label $\delta$, are observed in PIHS, but not in 
LTLS and HTHS.\cite{taya-PRL,okamura}    
In addition, the lines in 570-700~cm$^{-1}$ range show 
marked differences between PIHS and HTHS, except for 
the line $p4$ which is due to C-H deformation within the 
2-pic ligand.\cite{okamura}     
These lines have been attributed to skeletal vibrations of 
the aminomethyl 
group ($-$NH$-$CH$_2$$-$) in the 2-pic ligand.\cite{okamura}      
Hence the variations of these lines were regarded as evidence 
for microscopic deformations in the 2-pic ligand 
between HTHS and PIHS.

To analyze the data below 450~cm$^{-1}$, 
we have calculated the frequencies of infrared-active 
molecular vibrations for an isolated [Fe(2-pic)$_3]^{2+}$, 
using the density-functional method.   
The calculation was made using the Gaussian '03 
program,\cite{gaussian} the details of which were similar to 
those previously reported.\cite{juhasz}    
The parameters involved in the calculation were first optimized 
so as to reproduce the reported molecular structure of 
[Fe(2-pic)$_3]^{2+}$, then they were used to calculate the 
infrared frequencies.    The calculated vibration frequencies 
are displayed by the vertical bars in Figs.~4(a) and (b).    
It is seen that most of the observed lines have their 
frequencies close to the calculated ones.    
(Note that line $e$ should be due to ethanol, 
since it is hardly affected by the crossover.)       
The lines marked by the asterisks are located far apart 
from the calculated frequencies.   This deviation should 
be related with the strong hydrogen bonding between the 
amine group (-NH-) of 2-pic and Cl$^-$,\cite{mikami} 
which was neglected in the calculation.   
The calculated lines below 300~cm$^{-1}$ in the high-spin 
states and those below 400~cm$^{-1}$ in LTLS are 
mainly derived from the FeN$_6$ vibration modes.     
Hence the observed lines in these frequency ranges are 
also attributed to FeN$_6$-based vibrations (except for 
line $e$).    
The lines $p1$-$p3$ are attributed to intramolecular 
vibrations of 2-pic ligand, as in the previous Raman 
work.\cite{taya-PRL,taya-PRB}

In Fig.~4, the spectra below 300~cm$^{-1}$ in PIHS and HTHS 
[(b) and (c)] are very similar to each other.    
They mainly consist of three strong bands, which are labeled 
as $c$1-$c$3 in Fig.~4(c).    The spectral similarity demonstrates 
that the microscopic vibrational states of the FeN$_6$ 
cluster are also similar between PIHS and HTHS.   In contrast, 
the spectrum in LTLS, Fig.~4(a), appears quite different from 
those of the high-spin states.    The bands 
$c$1-$c$3 are no longer observed in LTLS.      
Instead, a larger number of narrower lines are observed, 
which implies a symmetry lowering of FeN$_6$ in LTLS.    
The spectral differences should be due to the deformation 
of FeN$_6$ and due to changes in the force constants of 
Fe-N bonds.   
The occupation of $t_{2g}$ 
orbitals by 6 electrons in LTLS, 
as sketched in Fig.~1(b), results in not only the shorter 
Fe-N distances, but also stronger Fe-N bonds.\cite{SC-review}     
Note that the symmetry of the FeN$_6$ in the average crystal 
structure is nearly the same between the high- and 
low-spin states.\cite{mikami,huby}    
Hence, the changes in the force constants seem to have 
lowered the symmetry of FeN$_6$ cluster mechanically 
(not geometrically), resulting in the appearance of a 
larger number of narrower lines in LTLS.

The present data have demonstrated that the microscopic 
vibrational states of FeN$_6$ are very similar between 
PIHS and HTHS.     This is consistent with the previous 
results of XRD,\cite{huby} XAFS\cite{oyanagi} and 
NRIS,\cite{juhasz} all of which gave very similar data 
between PIHS and HTHS.      
Compared with the previous works, however, it is very 
important that the present work has explicitly and 
directly shown the 
microscopic similarity with a high spectral resolution.     
In contrast, as already mentioned, the absorption 
spectrum at 530-700~cm$^{-1}$ range has shown clear 
differences between PIHS and HTHS, due to skeletal 
deformation of the 2-pic ligand.\cite{taya-PRB,okamura}    
Considering these results, therefore, {\it the most 
important microscopic difference between the PIHS and HTHS 
of Fe-pic should be the deformation of 2-pic ligands.}    
Note that this deformation does not have a long-range 
order, since the XRD data\cite{huby} show no appreciable 
deformation of 2-pic in the average crystal 
structure.

The unusual properties of PIHS, mentioned in the 
introduction, apparently result from a cooperative 
interaction (cooperativity) among Fe$^{2+}$ ions.     
However, a cooperativity is very important also in the 
thermal spin crossover between HTHS and LTLS.\cite{SC-review}   
This was experimentally demonstrated on the diluted system 
(Fe, Zn)-pic, where the spin crossover became much broader 
at low Fe fractions, approaching to that given by the 
Boltzmann distribution over isolated molecules.\cite{dilution}   
An important source of cooperativity is the long-range 
elastic interaction, caused by the deformation of FeN$_6$ 
upon the crossover.\cite{SC-review,dilution,elastic}     
In this mechanism, an increase in the density of 
high-spin Fe$^{2+}$ effectively increases the 
interaction (or equivalently lowers the energy 
difference between low- and high-spin states\cite{ogawa}), 
accelerating the crossover compared with the isolated 
case.\cite{dilution,ogawa}      
A theory based on the elastic interaction has also 
successfully reproduced two key properties of PIHS under 
photoexcitation,\cite{ogawa} i.e., the presence of 
incubation period and the threshold excitation 
intensity.\cite{koshihara}

In the above models, however, the microscopic properties 
involved in the interaction are not taken into account.      
In addition, short-range interaction among Fe$^{2+}$ ions 
is neglected.    It has been pointed out that the step-like 
change of $\gamma_{HS}$ and the phase separation in PIHS 
cannot be understood without the short-range 
interaction.\cite{ogawa}     It is therefore important 
to characterize the interaction among Fe$^{2+}$ ions 
more microscopically.    
As mentioned before, the -NH- portion of the aminomethyl 
group in the 2-pic ligand is strongly hydrogen-bonded 
to the Cl$^-$ anion.   This hydrogen bonding is also 
responsible for the intermolecular bonding and 
crystallization of Fe-pic molecules.\cite{mikami}       
Hence, the microscopic deformation of 2-pic ligand found 
in the present work is quite likely to reflect different 
states of intermolecular bondings between HTHS and PIHS.      
To further characterize such bonding, it should be very 
useful to study the vibration modes below the frequency 
range of this work.     
The intermolecular vibration modes, which involve 
the vibration of the entire [Fe(2-pic)$_3$]$^{2+}$ 
ion, are expected to appear well below 80~cm$^{-1}$.    
Such modes are expected to be more sensitive to 
changes in the intermolecular bonding than those 
observed in this work.

Low-frequency vibrations may be important also in terms 
of the vibrational entropy.\cite{entropy2}    
For the thermal crossover in Fe-pic, the phonon part 
($\Delta S_{ph}$) of 
the observed entropy change is as large as 
56~\%.\cite{entropy}     
This acts as a strong driving force for the crossover 
and also increases the cooperativity.\cite{SC-review,entropy3}   
The large $\Delta S_{ph}$ results from the strong 
anharmonicity of the FeN$_6$ vibrations: 
When the average lattice constants 
change upon the crossover, the phonon frequencies 
also change due to the anharmonicity.  
Consequently, the phonon density of states is modified, 
leading to the large $\Delta S_{ph}$.\cite{entropy2,entropy}  
Since the photoinduced crossover is observed at much lower 
temperatures, high-frequency phonons are quenched, and 
phonons with much lower frequencies may play important 
roles in terms of the entropy in PIHS.    It is interesting 
that, in Fig.~4(c), the absorption band $c1$ shows significant 
broadening in HTHS compared in PIHS.    This broadening 
of $c1$ seems unusually larger than those of $c2$ and $c3$, 
compared with phonons in the usual solids having similar 
frequencies.    This might be a sign of the strong 
anharmonicity of the low-frequency FeN$_6$ vibrations.    
Again, a further study at lower frequency range is 
needed to obtain more information about the role of phonon 
anharmonicity.

In conclusion, we have reported the first FIR absorption study 
of Fe-pic single crystals in its three characteristic states.  
The absorption lines below 400~cm$^{-1}$ are mainly attributed 
to the FeN$_6$ vibrations.      
The spectra are found very similar between PIHS and HTHS, 
which demonstrate that the microscopic environment at the 
FeN$_6$ cluster is also similar.    
The most important microscopic difference between 
HTHS and PIHS is the deformation of the 2-pic ligand, 
which should have important effects on the 
intermolecular coupling.     
The present result suggests the importance of further 
study at lower frequencies, which should give 
more insight into the microscopic nature of the 
intermolecular interaction in PIHS.   

The experiments at SPring-8 were performed under the 
approval of JASRI (2004A0480-NSa-np).


\pagebreak

\begin{figure}
\begin{center}
\includegraphics[width=0.5\textwidth]{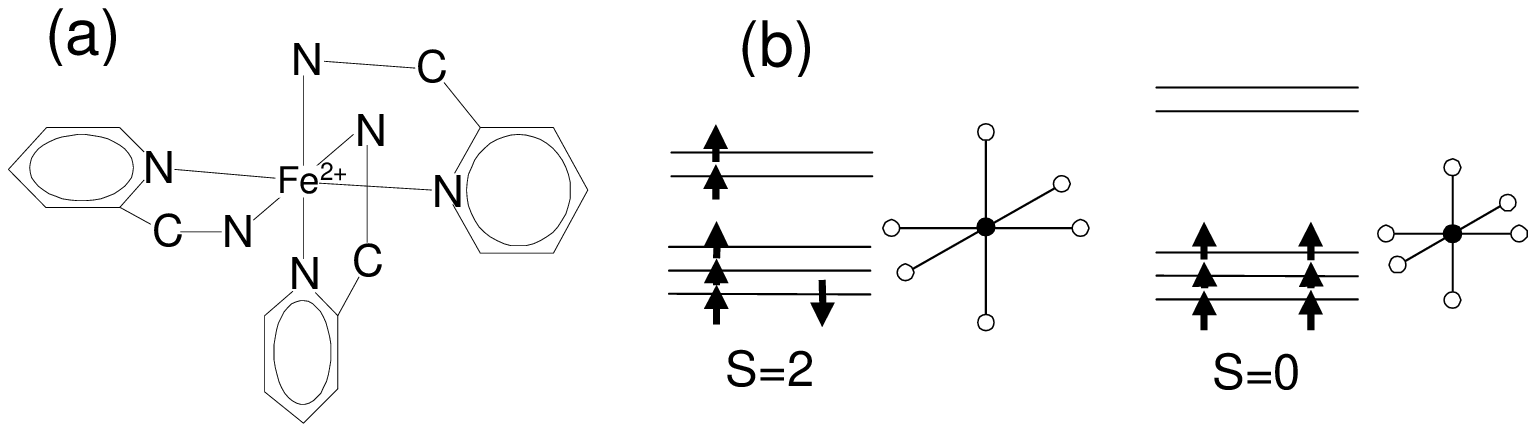}
\caption{
(a) The structure of [Fe(2-pic)$_3$]$^{2+}$.      
(b) Electron configurations of Fe$^{2+}$ in high-spin 
($S$=2) and low-spin ($S$=0) states.      
} 
\end{center}
\end{figure}

\begin{figure}
\begin{center}
\includegraphics[width=0.65\textwidth]{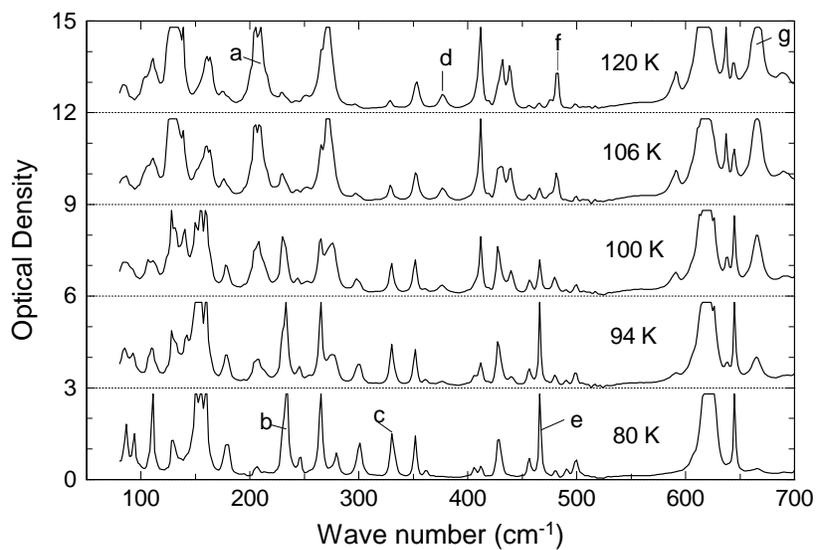}
\caption{
Optical density of Fe-pic at 
several temperatures without photo-excitation.    
The labels a$-$g indicate the absorption bands and 
lines analyzed in Fig.~3.   
}
\end{center}
\end{figure}

\begin{figure}
\begin{center}
\includegraphics[width=0.5\textwidth]{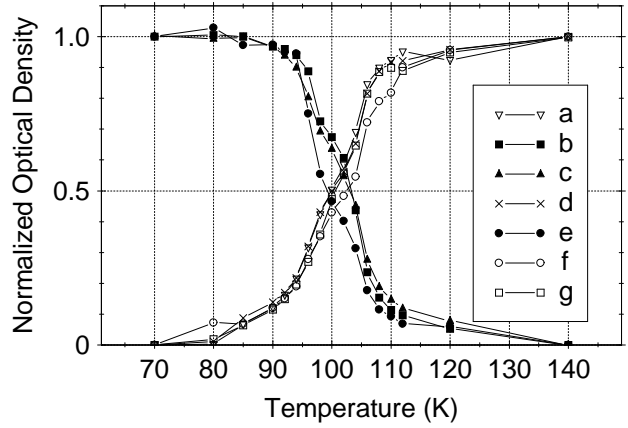}
\caption{
Optical density (OD) of selected absorption lines and 
bands for Fe-pic, which are indicated in Fig.~2 
by the labels a-f,  as a function of temperature.    
The OD has been integrated over wave number regions 
of (a) 187-224~cm$^{-1}$, (b) 224-241~cm$^{-1}$, 
(c) 322-340~cm$^{-1}$, (d) 365-392~cm$^{-1}$, 
(e) 462-471~cm$^{-1}$, (f) 473-488~cm$^{-1}$, 
and (g) 650-680~cm$^{-1}$, then normalized by 
the difference of OD between 70 and 140~K.     
} 
\end{center}
\end{figure}

\begin{figure}
\begin{center}
\includegraphics[width=0.7\textwidth]{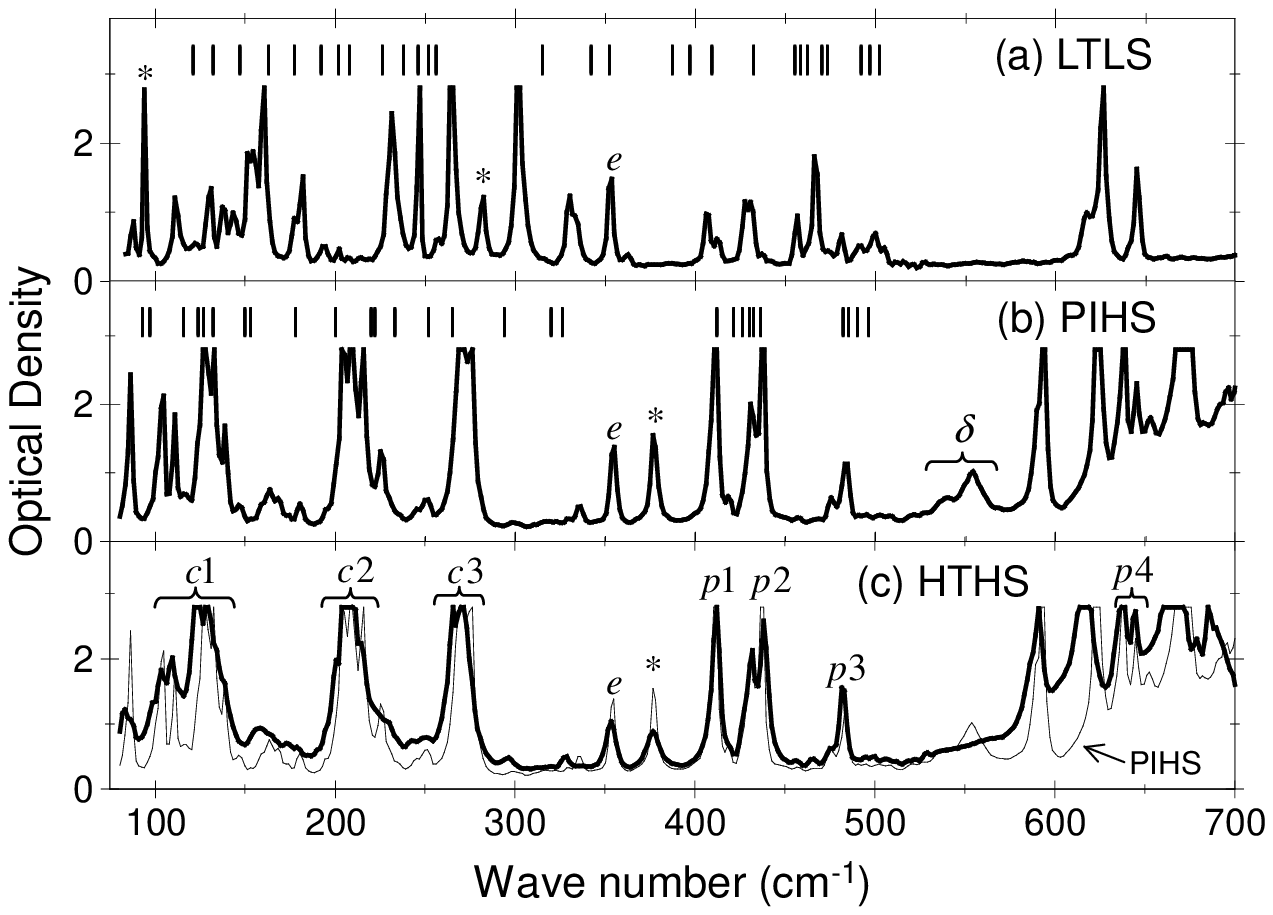}
\caption{
Optical density spectra of Fe-pic in 
(a) LTLS at 9~K, (b) PIHS at 9~K, and (c) HTHS at 140~K.    
The thin curve in (c) is the same as the spectrum in (b), 
shown for comparison.    
The vertical bars in (a) and (b) show the calculated 
frequencies of molecular vibrations for 
[Fe(2-pic)$_3$]$^{2+}$.    See text for the labels.  
}
\end{center}
\end{figure}


\begin{thebibliography}{99}
\bibitem{footnote4} 
The curves in Fig.~3 show the mid-point of the crossover 
around 103~K, which is lower than $T_{1/2}$.    
This difference is due to a low thermal contact of the 
sample with the cryostat, which resulted from our effort 
of not rigidly mounting the sample to avoid causing an 
internal strain.    
At lower temperatures, the thermal conductivity of the 
sample becomes much higher, and the difference between 
the measured and actual temperatures should be smaller.     
Our discussions below will not be affected by such 
temperature difference.
\bibitem{footnote3} 
Due to technical restrictions, the data in Fig.~3 were 
measured on a different sample from that for the data 
in Fig.~2.     The former was slightly thicker, resulting 
in the stronger absorption in Fig.~3.
\end{thebibliography}
\end{document}